\documentclass[letterpaper]{article}
\pdfoutput=1
\usepackage[top=3cm,bottom=3cm,left=3cm,right=3cm,marginparwidth=1.75cm]{geometry}
\usepackage{listings}
\usepackage{xcolor}
\usepackage{subfigure}
\usepackage{multirow}
\usepackage{dcolumn}
\usepackage[english]{babel}
\usepackage[utf8x]{inputenc}
\usepackage[T1]{fontenc}
\usepackage{palatino}
\usepackage{cite}
\usepackage{amsmath}
\usepackage{afterpage}
\usepackage{graphicx}
\usepackage[colorlinks=true, linkcolor=magenta, citecolor=magenta]{hyperref}

\begin{document}
\hfill{MIT-CTP/5199}

\vspace*{0.5cm}

\begin{center}

{\LARGE \bf Indirect Detection of Dark Matter in the Galaxy}

\par\vspace*{4mm}\par

{

\bigskip

\large \bf Rebecca K. Leane\footnote{Email: \href{mailto:rleane@mit.edu}{rleane@mit.edu}; ORCID: \href{http://orcid.org/0000-0002-1287-8780}{0000-0002-1287-8780}}}

\bigskip

{Center for Theoretical Physics \\
Massachusetts Institute of Technology \\
Cambridge, MA 02139, USA}

\bigskip

\end{center}

\begin{abstract}
I present a short overview of the latest developments in indirect searches for dark matter using gamma rays, X-rays, charged cosmic rays, micro waves, radio waves, and neutrinos. I briefly outline key past, present, and future experiments and their search strategies. In several searches there are exciting anomalies which could potentially be emerging dark matter signals. I discuss these anomalous signals, and some future prospects to determine their origins.
\end{abstract}

\begingroup
\hypersetup{linkcolor=black}
\tableofcontents
\endgroup

\section{Introduction}

Indirect searches for dark matter are incredibly exciting. The Universe has been exploding systems, smashing systems together, and allowing systems to decay since the beginning of time. This provides an enormous advantage in searches for dark matter compared to collider and direct detection efforts -- we can access unique lengths and energies using experiments the Universe has been running \textit{for us} over very long time scales. Compared to Earth-based searches, we can probe higher energies, longer particle decay lengths, and weaker particle couplings. Furthermore, we can observe dark matter in its natural habitat. All the evidence to date for dark matter comes from its fingerprints in astrophysics and cosmology. The defining feature of thermal particle dark matter, its annihilation cross section, can be directly compared to its annihilation rate observed in indirect experiments. This provides a clear target for dark matter discovery, or exclusion of particular models.

In this mini-review, I provide a brief overview of searches for dark matter using gamma rays, cosmic rays, and neutrinos. I will discuss the latest developments for hints for dark matter in these observables. I detail future paths enabling the fundamental particle nature of dark matter to potentially be finally revealed.

\section{Ingredients for Indirect Dark Matter Searches}

Indirect dark matter searches scan the sky for any excess Standard Model (SM) particles or anti-particles produced from dark matter annihilation or decay. The flux $\Phi$ of neutral particles\footnote{Note that for charged particles arising from dark matter annihilation or decay, their flux will contain additional terms relevant for cosmic ray propagation, see e.g. Ref~\cite{Amato:2017dbs}. Also note that redshift (or absorption) effects are not shown here, which can be relevant when particles traverse large distances/times, see e.g. Ref.~\cite{Slatyer:2015jla}.} arising from the annihilation ($a=2$) or decay ($a=1$) of dark matter with mass $m_\chi$ can be described as
\begin{equation}
    \Phi(E,\phi)= 
    \frac{\Gamma}{4\pi m_\chi^a}\frac{dN}{dE}\int \rho[r,(\ell,\phi)]^a \, d\ell.
\end{equation}
This equation encapsulates three important aspects for the dark matter sourced particle rate:
\begin{enumerate}
 \item The interaction rate of dark matter particles $\Gamma$. For annihilating dark matter, the rate is $\Gamma=\langle\sigma v\rangle/2$, where $\sigma$ is the cross section for dark matter particles with relative velocity $v$. For dark matter decay, the interaction rate is $\Gamma=1/\tau$, where $\tau$ is the dark matter lifetime.
 \item The energy spectrum of the annihilating/decaying particles ($dN/dE$). Depending on the mass of the dark matter particles, different amounts of energy can be imparted to the detectable annihilation/decay products. Further, depending on what the products are, their energy spectrum will vary depending on how/if they hadronize, how/if they decay, and how/if they radiate any particles themselves.
 \item A piece arising from astrophysics, which relates to the dark matter density in the relevant environment. This factor is the integral over the dark matter density $\rho$ raised to some power $a$ along the line of sight $d\ell$. For annihilating dark matter, this is called the J-factor~\cite{Bergstrom:1997fj}, with $a=2$. For decaying dark matter, this is called the D-factor~\cite{Geringer-Sameth:2014yza}, with $a=1$. This relies on information from $N$-body simulations, which informs of the possible dark matter density profiles present in the system of interest.
\end{enumerate}
While indirect searches for dark matter are advantageous over other searches in many ways, they have a significant difficulty: not-well understood backgrounds. Along with this, can come large systematic errors. To optimize discovery or exclusion, we want to search in a way that maximizes signal over background (or minimizes potential unknown systematics). To do this, we want to exploit the relevant inputs to the dark matter particle flux listed above. This can be optimized by:
\begin{enumerate}
 \item Looking for scenarios (or times) where the annihilation or decay rate is enhanced. For example, Sommerfeld enhanced annihilation can lead to larger rates today than in the early Universe. Importantly however, for many dark matter models, the annihilation rate will be set by the relic density of dark matter in the Universe today, and as such may not be able to be freely increased or decreased without consequence. 
 \item Looking in energy bands where the signal energy spectrum is distinct or peaked relative to the shape of the background spectrum. This can happen for example where line or box spectral features arise.
 \item Looking in places where the dark matter density is large. A good example here is the galactic center.
\end{enumerate}
Equipped with these points, lets see what we can find throughout the Universe, and what has been potentially found so far!

\section{Dark Matter Searches with Neutrinos}
\label{sec:neutrinos}

\subsection{Experiments and Prospects}

Searches with neutrinos provide unique opportunities and challenges. The very weak interactions of these ghostly particles provide a window into the deepest and darkest places in the Universe, where no other signals can escape. On the other hand, exactly because they are so weakly interacting, they can then be difficult to detect.

\begin{figure}[t]
\centering
\includegraphics[width=0.72\textwidth]{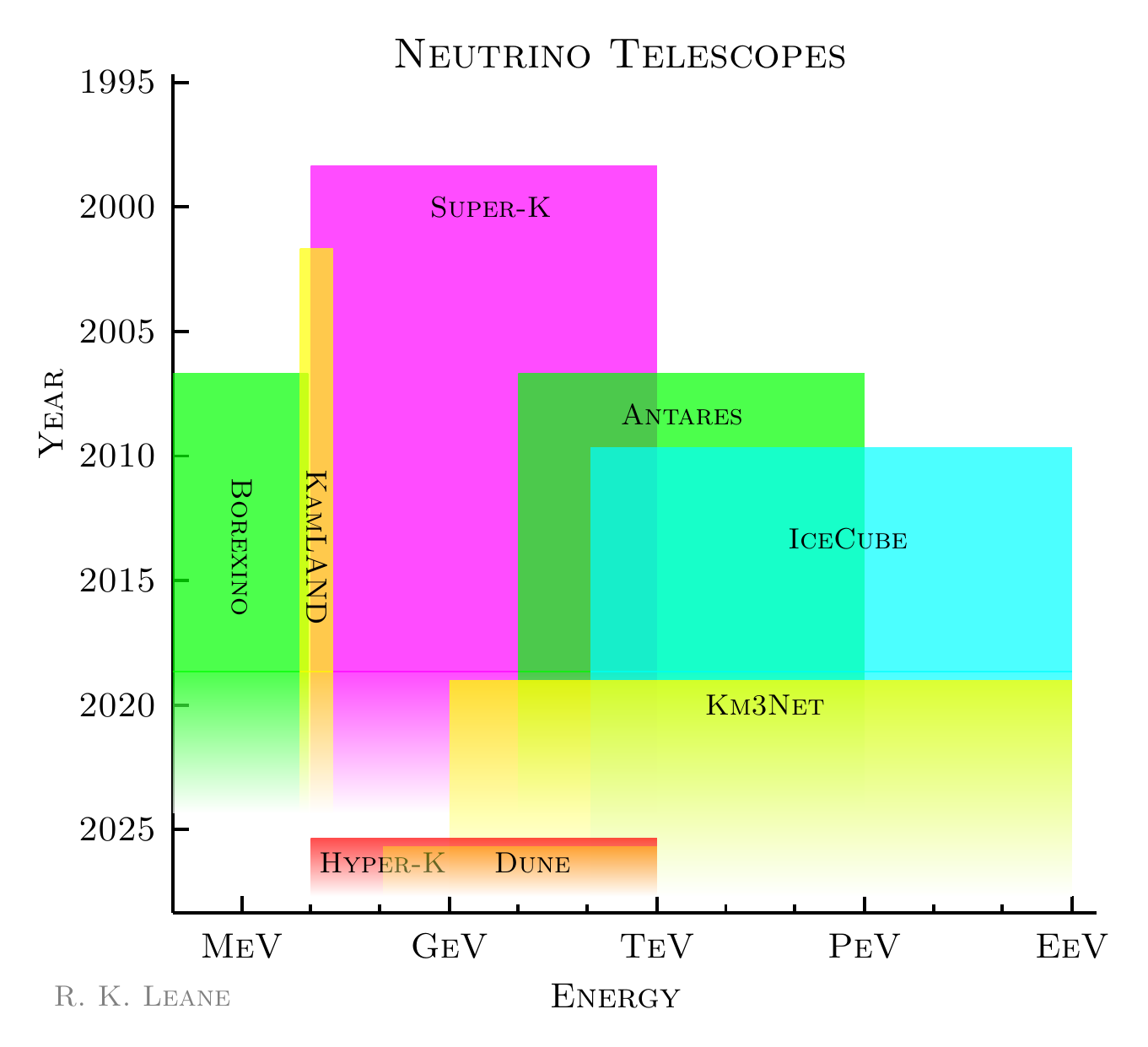}
\caption{Summary of selected neutrino experiments directly searching for dark matter annihilation, by mission dates and approximate energy sensitivity.}
\label{fig:neutrinos}
\end{figure}

Figure~\ref{fig:neutrinos} shows a brief overview of selected neutrino experiments searching for dark matter in the last couple of decades, as a function of their approximate energy sensitivity. Just in 2019, KM3Net has come online, and is beginning to provide preliminary results on dark matter annihilation~\cite{km3netsun,gozzini}. IceCube-Gen2 (the IceCube upgrade) will substantially improve sensitivity to dark matter masses below around 100 GeV for high-energy solar neutrino searches, by many orders of magnitude~\cite{In:2017kcf}. Coming soon is Super-K's successor, Hyper-K. The fiducial volume of the Hyper-K tank is about 10 times larger than the Super-K tank, leading to a large improvement on previous flux sensitivities.  This will provide better sensitivities to the dark matter annihilation cross section by about an order of magnitude~\cite{Abe:2018uyc,Arguelles:2019ouk,Bell:2020rkw}. For solar neutrino searches, Borexino has sensitivity to $\sim$MeV mass annihilating dark matter ~\cite{Bellini:2010gn}. Limits on $\sim$MeV mass dark matter can also be set using the flux of extraterrestial neutrinos with KamLAND~\cite{Collaboration:2011jza,Arguelles:2019ouk}. Dune is a new experiment currently under construction, and compared to Cherenkov detectors such as Super-K, IceCube, and ANTARES, has improved energy resolution to observe neutrinos from dark matter annihilation~\cite{Acciarri:2015uup,Capozzi:2018dat}. Dune can be used in future for several dark matter-neutrino related contexts; behind Hyper-K, it may provide the next best sensitivity to galactic dark matter for $\sim$GeV dark matter masses, down to annihilation cross sections of $\sim10^{-24}$cm$^3$/s~\cite{Arguelles:2019ouk}.

\subsection{Dark Matter Annihilation and Decay into Neutrinos}

The current leading limits on dark matter annihilation into neutrinos, for most dark matter masses, come from observations of the galactic center, where locally the dark matter density is greatest. However, compared to searches with the same targets using with gamma or cosmic rays, these limits are generally the weakest (and certainly above the thermal relic cross section), as neutrinos are much more difficult to detect. 

\begin{figure}[h!]
\centering
\includegraphics[width=0.86\textwidth]{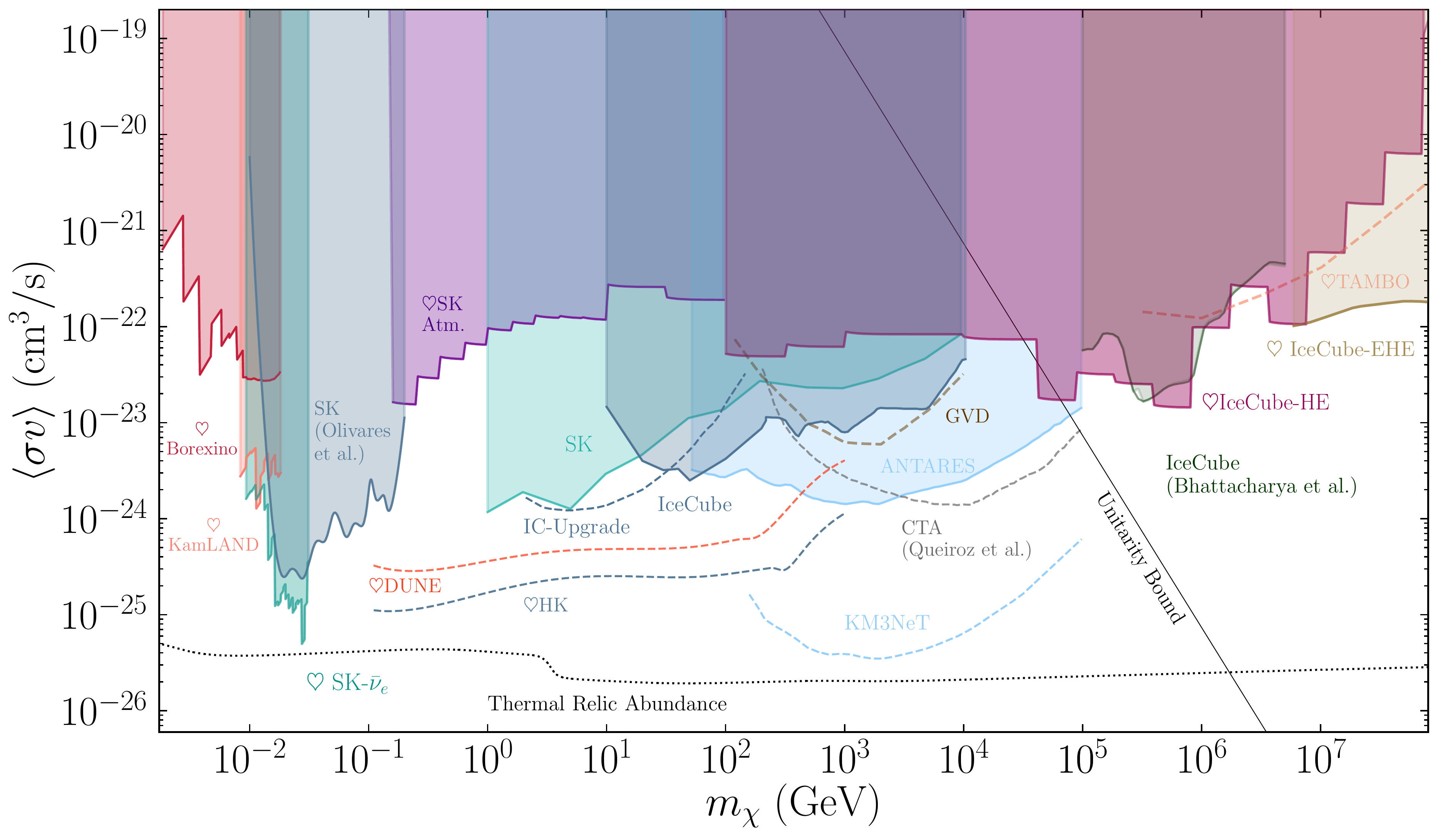}
\caption{Summary of 90$\%$ C.L. limits (shaded) and projected sensitivities (dashed) to the dark matter annihilation cross section, assuming $s$-wave annihilation. Plot taken from Ref.~\cite{Arguelles:2019ouk}.}
\label{fig:neutrinoplot}
\end{figure}

Figure~\ref{fig:neutrinoplot} shows a recent summary of the landscape of constraints on the dark matter annihilation cross section from neutrinos. See Ref.~\cite{Arguelles:2019ouk} for further details.


The diffuse neutrino flux in the Milky Way halo can also be used to search for dark matter~\cite{Beacom:2006tt,Yuksel:2007ac}. In fact, a hint of dark matter has been suggested to potentially exist in IceCube's observed TeV-PeV diffuse neutrino flux~\cite{Aartsen:2013jdh}. The origin of such high-energy neutrinos is not yet understood, so has been purported to potentially contain a component from decaying dark matter (see e.g.~\cite{Esmaili:2013gha,Feldstein:2013kka}). However, this is constrained in part by gamma-ray observations, as decaying dark matter that produces neutrinos should also produce gamma rays -- and no such excess gamma rays are observed by the gamma-ray telescope Fermi~\cite{Cohen:2016uyg,Chianese:2019kyl}. Therefore, the current status is that dark matter could still contribute to the diffuse TeV-PeV neutrino flux, though parameters which could give rise to this flux are partly constrained.

\subsection{Dark Matter Scattering and Annihilation to Neutrinos in the Sun}

\begin{figure*}
     \begin{center}
        \subfigure{
            \includegraphics[width=0.412\textwidth]{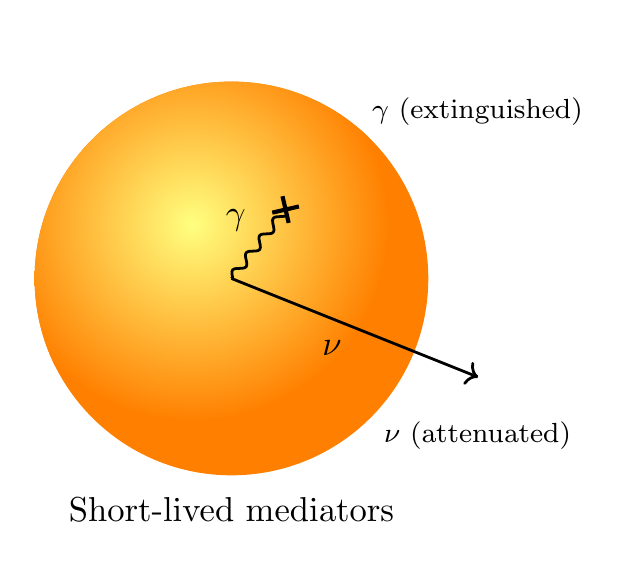}}
        \subfigure{
           \includegraphics[width=0.412\textwidth]{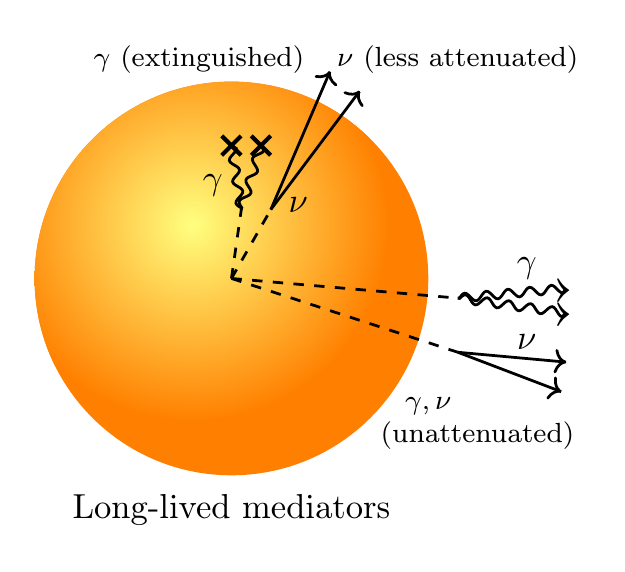}}
    \end{center}
    \caption{{\bf Left:} Short-lived mediator scenario. Only neutrinos can escape the Sun and they are attenuated. {\bf Right:} Long-lived dark mediator scenario. Gamma rays (and other particles) can escape, and neutrinos are less attenuated. Figure taken from Ref.~\cite{Leane:2017vag}.}
   \label{fig:sun}
\end{figure*}

Due to their ghostly nature, one of the best things about using neutrinos is peering into environments in which no other particles could escape. Such a way to search with neutrinos is to look for neutrinos escaping stars, like the Sun. Dark matter particles sweeping through the Sun can scatter with the solar matter, lose energy, and become gravitationally captured. Dark matter can then accumulate at the core, and if present with anti-particles, can annihilate to high-energy neutrinos which may escape the Sun and be detected on Earth~\cite{Press:1985ug,Krauss:1985ks,PhysRevLett.55.257,Peter:2009mk}. This provides a complementary probe of both the dark matter scattering cross section (as this determines how much dark matter was captured), as well as the dark matter annihilation cross section. In the scenario that dark matter annihilates to long-lived mediators, the neutrino flux is boosted as the particles are not as attenuated~\cite{Meade:2009mu,Bell:2011sn} (and in such a scenario, other particles can escape, such as gamma rays or charged particles~\cite{Batell:2009zp, Schuster:2009au, Schuster:2009fc, Meade:2009mu,Feng:2015hja,Feng:2016ijc,Leane:2017vag,Nisa:2019mpb,Albert:2018vcq,Albert:2018jwh,Niblaeus:2019gjk}). 

Figure~\ref{fig:sun} shows a comparison of these processes in the Sun. For similar processes but for dark matter in other stars such as neutron stars and white dwarfs, see Refs.~\cite{Goldman:1989nd,
Gould:1989gw,
Kouvaris:2007ay,
Bertone:2007ae,
deLavallaz:2010wp,
Kouvaris:2010vv,
McDermott:2011jp,
Kouvaris:2011fi,
Guver:2012ba,
Bramante:2013hn,
Bell:2013xk,
Bramante:2013nma,
Bertoni:2013bsa,
Kouvaris:2010jy,
McCullough:2010ai,
Perez-Garcia:2014dra,
Bramante:2015cua,
Graham:2015apa,
Cermeno:2016olb,
Graham:2018efk,
Acevedo:2019gre,
Janish:2019nkk,
Krall:2017xij,
McKeen:2018xwc,
Baryakhtar:2017dbj,
Raj:2017wrv,
Bell:2018pkk,
Garani:2018kkd,
Chen:2018ohx,
Garani:2018kkd,
Hamaguchi:2019oev,
Camargo:2019wou,
Bell:2019pyc,
Acevedo:2019agu,
Garani:2019fpa,
Joglekar:2019vzy,
Joglekar:2020liw}.

\section{Dark Matter Searches with X-Rays}

\subsection{Experiments and Prospects}

\begin{figure}[b!]
\centering
\includegraphics[width=0.45\textwidth]{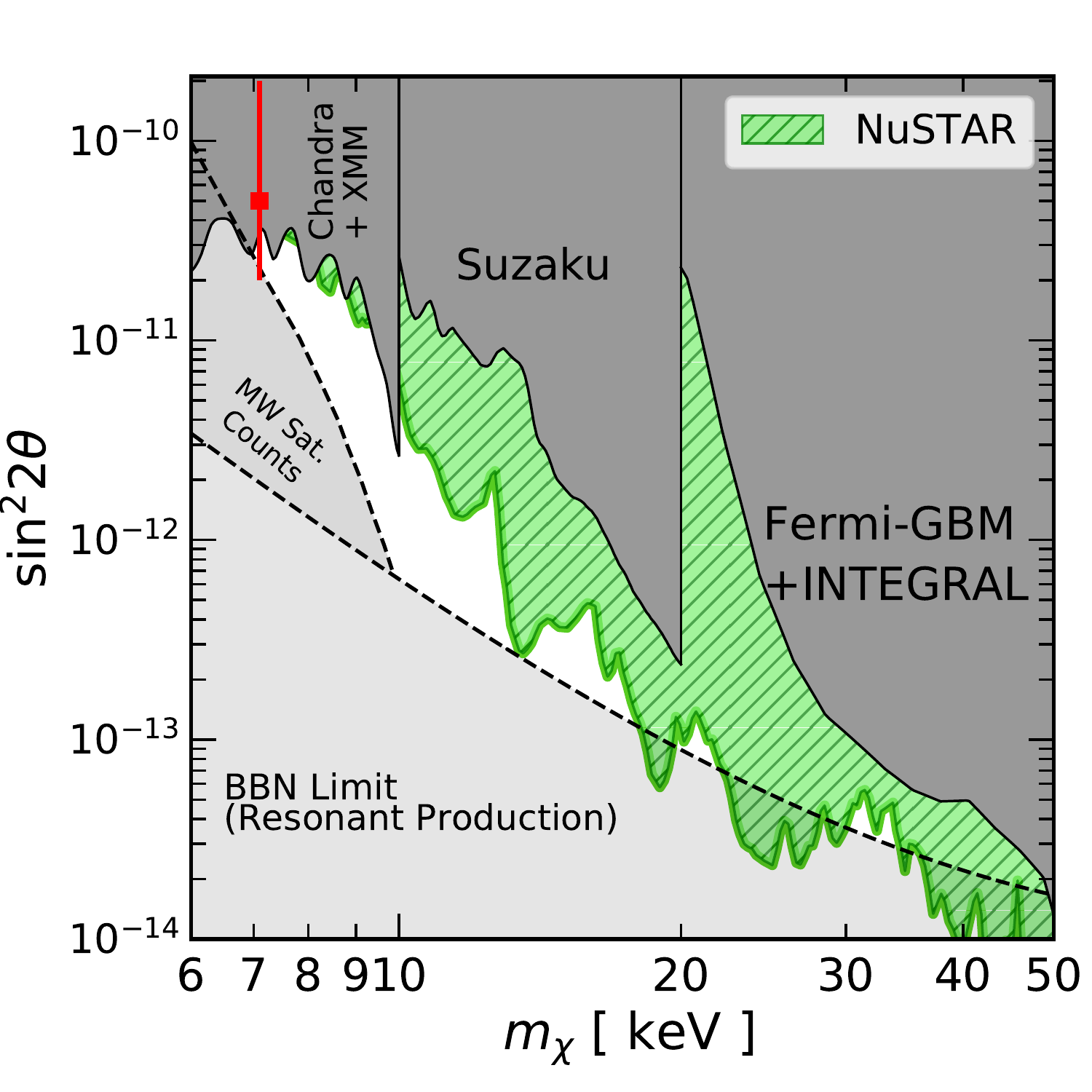}
\hspace{2mm}
\includegraphics[width=0.49\textwidth]{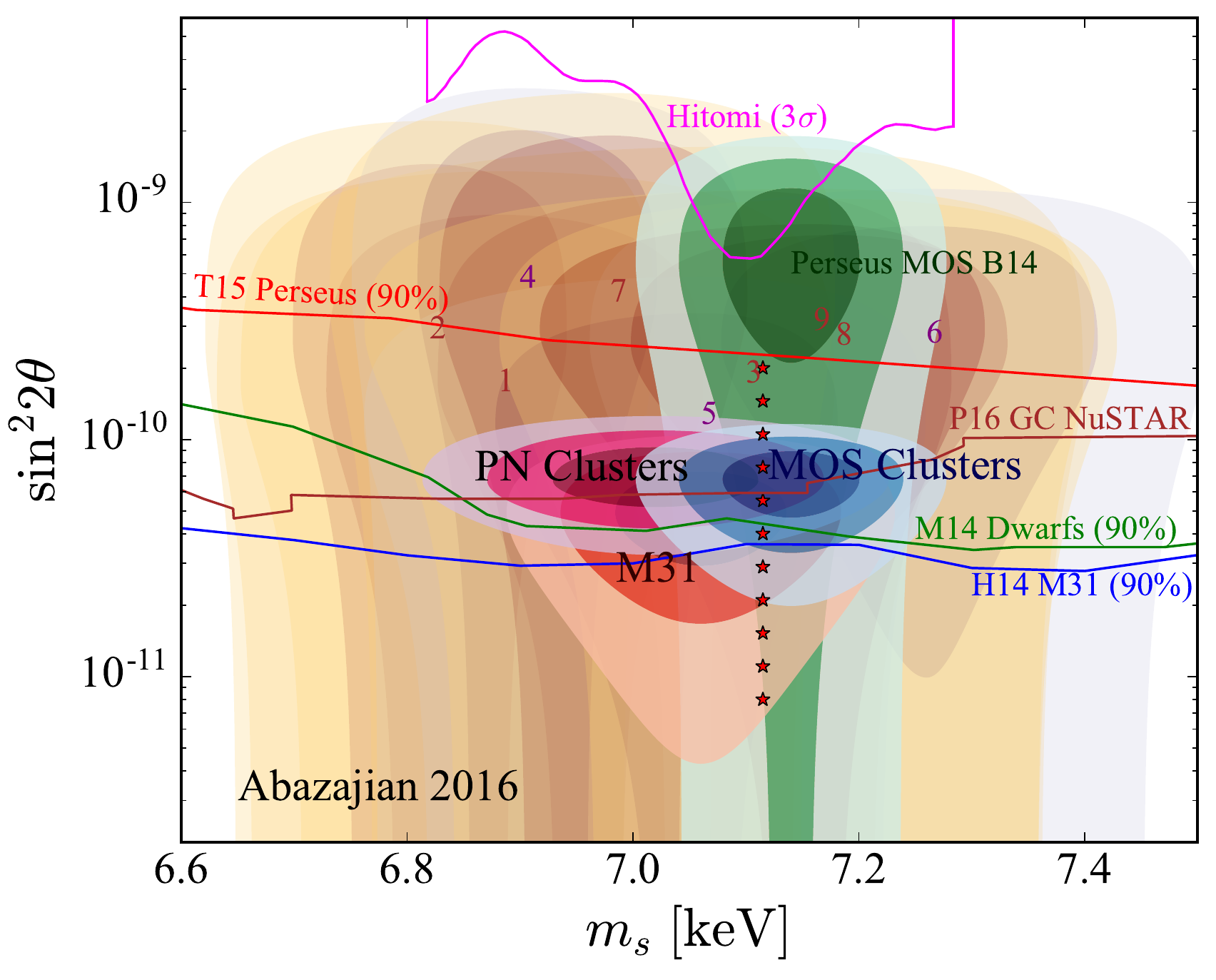}
\caption{\textbf{Left:} Summary of X-ray constraints on the sterile neutrino parameter space, taken from Ref.~\cite{Roach:2019ctw}. The red data point corresponds to an explanation of the 3.5 keV line. \textbf{Right:} Summary of best-fits (and some constraints) to the 3.5 keV line for the sterile neutrino model, taken from Ref.~\cite{Abazajian:2017tcc}. Ref.~\cite{Dessert:2018qih} argues for an additional upper constraint at sin$^22\theta\sim10^{-12}-10^{-11}$ in this mass range, which would present strong tension with the signal as dark matter, though there is some debate (see text).}
\label{fig:35kev}
\end{figure}

Like neutrinos, a benefit of X-ray signatures is that, being neutral particles, they are not subject to turbulent astrophysical magnetic fields, and propagate otherwise directly from their source (if not absorbed). Electrically charged dark matter annihilation products can lead to X-ray signatures, through inverse Compton scattering with astrophysical magnetic fields. Alternatively, the X-ray band provides an interesting probe of a keV-mass dark matter candidate, the sterile neutrino. In this case, sterile neutrino can decay into an active neutrino and photon, leading to an X-ray line (see below). 

Figure~\ref{fig:gammaxrays} shows a range of experiments searching for dark matter signals in X-ray (and gamma rays, see section below), as an approximate function of energy sensitivity and mission times.

\subsection{3.5 keV line}

Perhaps the most striking potential dark matter hint in X-ray at the moment is the 3.5 keV line. This may arise from a process $\chi\rightarrow\nu+\gamma$, where $\chi$ is a 7 keV sterile neutrino decaying to an active neutrino $\nu$ and photon $\gamma$. This line was first detected in stacked galaxy clusters with XMM-Newton~\cite{Bulbul:2014sua,Boyarsky:2014jta}. Intriguingly, it has been shown to be consistent with many complementary observations, though there is at least a slight tension with some other constraints~\cite{Abazajian:2017tcc}. 

Figure~\ref{fig:35kev} shows an overview of X-ray constraints on sterile neutrinos; see Refs.~\cite{Perez:2016tcq,Ng:2019gch,Roach:2019ctw} for more details on the left figure, and Ref.~\cite{Abazajian:2017tcc} for more details on the right figure.

It has been recently argued that using blank-sky searches, no 3.5 keV line is detected, allowing a strong constraint to be set on the sterile neutrino dark matter explanation~\cite{Dessert:2018qih}. This is because dark matter should be present everywhere in the halo (including the otherwise ``blank sky''), and so the line should have been detected if it were coming from decaying dark matter. However, study of a larger albeit overlapping region by other authors claim the line is indeed found~\cite{Boyarsky:2018ktr}. The discrepancy is claimed to arise due to too small an energy window used in Ref.~\cite{Dessert:2018qih}, and claims that other lines and instrument features not being correctly modeled~\cite{Boyarsky:2018ktr,Boyarsky:2020hqb,Abazajian:2020unr}. Considering the larger region of study~\cite{Boyarsky:2018ktr,Boyarsky:2020hqb,Abazajian:2020unr} which includes other background X-ray lines, the background is more supported and can result in its normalization being pushed down, such that a signal is visible. This is not a feature of the smaller energy window. However, it is interesting to note that in Ref.~\cite{Dessert:2018qih}, there does appear to be a large downwards fluctuation of the background in the signal region, such that the data is much lower than the expected background, leading to very strong limits. The treatment of the energy window and backgrounds appears to be the main point that is debated, and leads to bounds on sin$^22\theta$ that vary by a factor of $\sim8$.
 
It is also argued that including additional systematics from uncertainty in the dark matter density profile weakens the limits of Ref.~\cite{Dessert:2018qih} by an additional factor of $\sim3$~\cite{Abazajian:2020unr}. This is certainly true, although changing the dark matter profile and associated uncertainties would likely move the parameters corresponding to a signal detection from other Milky Way observations as well.

The origin of this line would have been settled in 2016 by satellite Hitomi, which unfortunately was destroyed only weeks after launching. New telescopes will be required to settle the debate once and for all. This should be possible within the next decade or so, with several X-ray telescopes and observatories to be launched which should be sensitive to an anomalous 3.5 keV line, such as e.g. Micro-X, X-Prism, and Athena. See Ref.~\cite{Boyarsky:2018tvu} for more details.

\section{Dark Matter Searches with Gamma Rays}
\label{sec:gammarays}

\subsection{Experiments and Prospects}

Dark matter annihilation can yield gamma rays, by hadronization of the final states, radiating gamma rays, or annihilating directly into pairs of gamma rays (producing a gamma-ray line). Generally, any dark matter model which has hadronic final states, will produce a strong signal in gamma rays. We currently search for dark matter in gamma rays from the Sun, the center of our galaxy, the Milky Way halo, in other galaxies, and in extragalactic environments. Like X-rays and neutrinos, gamma rays also benefit from not being subject to astrophysical magnetic fields, and if not absorbed, propagate directly from their source.

\begin{figure}[t]
\centering
\includegraphics[width=0.8\textwidth]{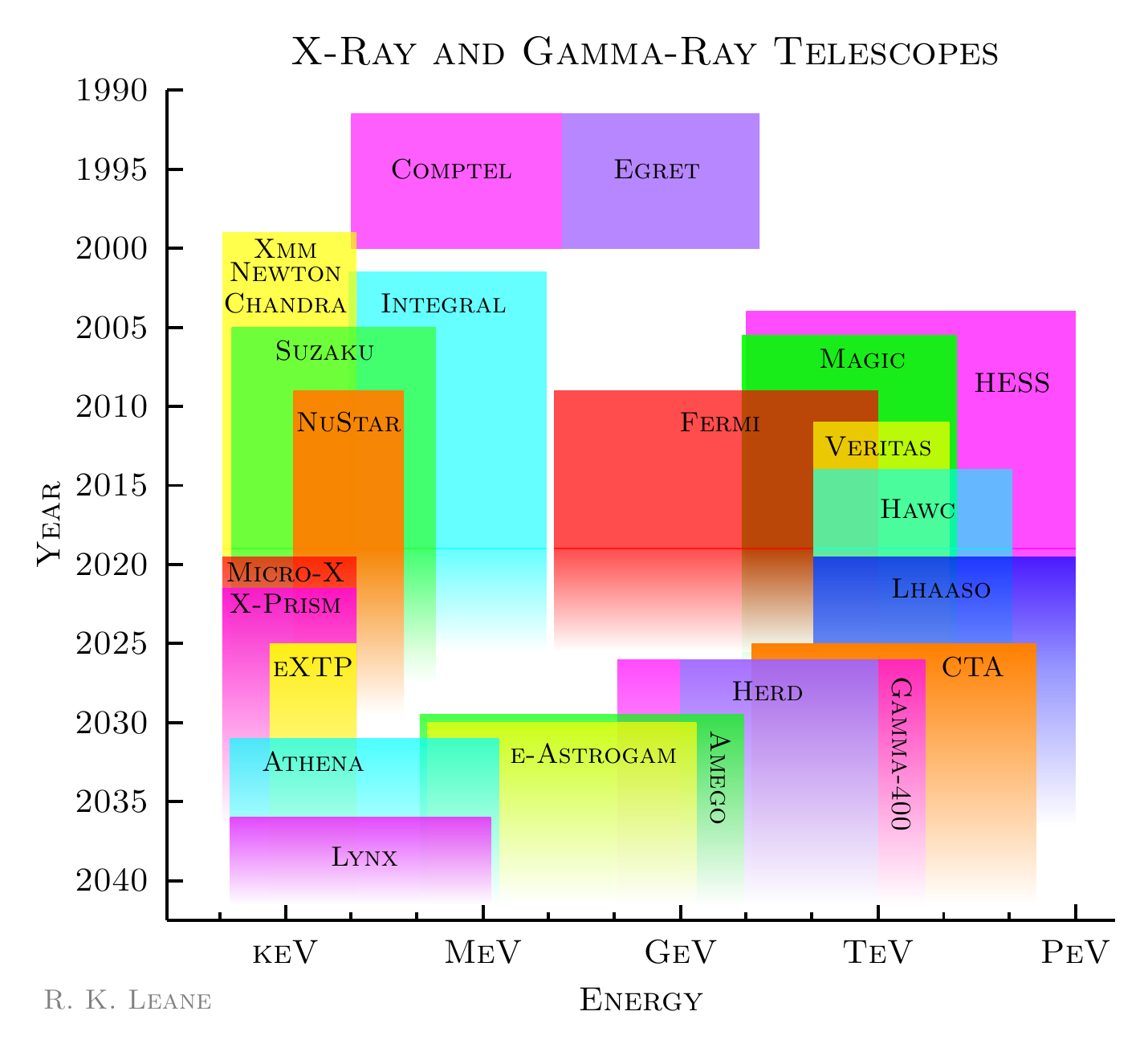}
\caption{Summary of selected X-ray and gamma-ray experiments by mission dates, and approximate energy sensitivity.}
\label{fig:gammaxrays}
\end{figure}

Figure~\ref{fig:gammaxrays} shows a range of experiments searching for dark matter signals in gamma ray (and X-rays, see section above), as a function of the approximate energies they are sensitive to, and their mission times in years. Our searches have been revolutionized by the Fermi Gamma-Ray Telescope (usually referred to as just ``Fermi''), which provides leading sensitivity to nearly all GeV gamma-ray targets. It has now collected over 10 years of data, with excellent angular and energy resolution in the GeV range. Above 1 TeV, other telescopes and observatories are superior, with progress rapidly improving. Currently, experiments such as MAGIC, VERITAS, HESS, HAWC, and LHAASO are providing sensitivity to the dark matter annihilation cross section. MAGIC, VERITAS and HESS are Imaging Atmospheric Cherenkov Telescopes (IACTs), and cannot look directly at nearby high-energy sources such as the Sun. They are ideal for e.g. TeV-gamma-ray Galactic Center searches. HAWC and LHAASO are water Cherenkov telescopes, and detect TeV gamma rays created in air-showers in the atmosphere (i.e., they don't measure the original cosmic rays/gamma rays directly). This means they are able to observe bright nearby sources such as the Sun, though they are also able to perform other searches (e.g. TeV-PeV gamma rays from the Galactic Center, dwarf spheroidals, etc).  We can expect exciting developments soon with CTA (Cherenkov Telescope Array). CTA is a set of over 100 ground-based telescopes, collectively using large mirrors and high-speed cameras to detect the cherenkov light produced by charged cosmic rays and gamma rays in atmospheric air showers. We can be excited for a significant boost in dark matter search sensitivity soon with CTA, which should extend below the thermal relic cross section at TeV masses~\cite{Pierre:2014tra,Silverwood:2014yza}.

\subsection{Galactic Center}

\subsubsection{GeV Excess}

The GeV Excess is an anomalous flux of GeV gamma rays peaked at $\sim2-3$ GeV, detected by the Fermi Gamma-Ray Space Telescope. It was first discovered in 2009~\cite{Goodenough:2009gk}, and has led to considerable debate as to its origin ever since. It is quite an exciting anomaly, as it first presented with many features of thermal particle dark matter signal. It features:
\begin{itemize}
 \item Consistent intensity of a thermal WIMP: the annihilation cross section required to fit the signal is $\langle\sigma v\rangle\sim 10^{-26}$cm$^3$/s~\cite{Daylan:2014rsa,Calore:2014xka}, which is of the same order obtained in the standard freezeout scenario.
 \item Consistent energy spectrum of a dark matter particle with mass $\sim50-100$ GeV~\cite{Daylan:2014rsa,Calore:2014xka}, best fit if the annihilation products are hadrons. Furthermore, the energy spectrum appears to be consistent throughout the excess, rather than changing with spatial distance from the center. This implies whatever is creating the excess is likely locally producing it, rather than it coming from some outflow event from the galactic center.
 \item Consistent morphology: it can be well fit with an NFW$^2$ dark matter profile~\cite{Daylan:2014rsa,Calore:2014xka} (though note that this point has since been debated\footnote{More recently, studies have shown that the morphology may be more like the stellar bulge than NFW$^2$~\cite{Macias:2016nev,Bartels:2017vsx,Macias:2019omb,Abazajian:2020tww}. Indeed, this would provide strong evidence against the dark matter explanation. However, this has been shown in some analyses to be dependent on modeling choices~\cite{Macias:2016nev,Bartels:2017vsx}, especially relating to the spatial tails of the excess, where the dark matter and bulge morphologies are most different, and the signal is over background is not large. Furthermore, given even the best current background models lead to very poor chi-squared values, it is not clear if systematic effects have yet been fully bracketed.}), and extends well out of the center of the galaxy, to at least 10 degrees. Visible matter, on the other hand, mostly correlates with the disk.
\end{itemize}

While the signal appears to nicely fit with the dark matter origin, there is a leading alternative: pulsars. Pulsars are rapidly spinning neutron stars. Those with millisecond periods, called millisecond pulsars, appear to be the leading candidate. This is mostly because the millisecond pulsars observed in other environments (specifically globular clusters) have a gamma-ray energy spectrum that also appears to approximately match the GCE~\cite{Cholis:2014noa,Cholis:2014lta}. 

Now, while pulsars are a good candidate for the excess, we still haven't seen any of the pulsars that are contributing GCE (if they are), which is argued to be surprising given the number of low-mass X-ray binaries observed in the GC~\cite{Cholis:2014lta,Haggard:2017lyq}. Furthermore, there are questions of how it is possible to have the number and distribution of pulsars that are required to explain the excess~\cite{Cholis:2014lta,Zhong:2019ycb}.

The general point is that there are two good leading explanations, and it is difficult to know which is the correct answer. Knowing the answer, however, is of pressing importance. We know dark matter makes up most of the matter in the Universe, yet we still have no information as to its fundamental particle nature. If the GCE turns out to be dark matter, it would be the first evidence of dark matter interacting with the Standard Model, and would have dramatic implications for all of physics.

So let's try to answer the question: what is the best explanation of the data, dark matter, or pulsars? To try to answer this question by searching for evidence for point sources (note pulsars appear as point sources to Fermi), there are two main methods:

\begin{itemize}
 \item Non-Poissonian Template Fitting (NPTF): This method exploits the difference in photon statistics expected from dark matter vs point sources. Specifically, gamma rays from dark matter are expected to provide smooth diffuse glow. This is because it is spread approximately smoothly through a halo. Gamma rays from point sources, however, look more clumpy. You can imagine a given pixel has no point sources, while another pixel has several, leading to more gamma rays. The variance across pixels in the sky can be much higher, leading to the clumpy image. 
 We can then build up a picture of the gamma-ray sky by modeling individual spatial components which contribute in gamma rays, called ``templates'', and assign them either the non-Poissonian (clumpy) statistics, or Poisson (smooth) statistics. These templates are then all floated in a fit, which aims to reveal if the preference is for the GCE to look more dark matter like (smooth), or point source like (clumpy)~\cite{Lee:2014mza,Lee:2015fea}.
 
 \item Wavelets: This method uses a wavelet transform to look for peaks in the data, which could be attributed to point sources. A smooth GCE such as that from dark matter would not be expected to produce peaks. If enough peaks are found with cumulative intensity of the GCE, this would be evidence for the point source origin~\cite{Bartels:2015aea}.
\end{itemize}

In 2015, evidence for clumpy rather than smooth GCE signals was independently found using both these techniques by independent groups~\cite{Lee:2015fea,Bartels:2015aea}. This was a seemingly fatal blow to the dark matter interpretation, and had led to the community consensus that the GCE was instead likely due to point sources.\\

\textbf{2019 Updates}\\

2019 presented a \textit{double} plot twist for these methods, with the interpretation of both the independent 2015 results challenged by two independent groups.

In Ref.~\cite{Leane:2019xiy}, using the NPTF, it was shown that mismodeling of the templates can hide a dark matter signal. In a proof-of-principle scenario, where an unmodeled set of point sources were simulated alongside a dark matter signal, the dark matter signal was misidentified as a new population of GCE point sources. Most importantly, evidence of some mismodeling effect was found in the real data. When injecting a dark matter signal into the data, the dark matter signal was not recovered, and instead was misattributed to the GCE point source template in the fit. Perhaps worse, when allowing the dark matter template to take negative values, the fit preferred deeply negative flux values, which clearly is not physical. The degree in which the dark matter was driven negative varied with Galactic Galactic diffuse emission models, and so it appeared likely this behavior was driven by mismodeling of the poorly-understood Galactic diffuse emission model\footnote{Galactic diffuse gamma rays are the largest contribution to the gamma-ray sky, and arise from cosmic rays interacting with the gas and starlight in our Galaxy. Modeling for the diffuse gamma rays is not well understood, and is often the largest source of uncertainty in gamma ray searches.}. More broadly, this paper argued that given that systematics were clearly not under control, no robust evidence using the NPTF could be claimed for (or against) the point source interpretation of the data.

In Ref.~\cite{Zhong:2019ycb}, the interpretation of the wavelet method results were challenged. It was shown that when updating the wavelet analysis to mask out Fermi's new point source catalog (4FGL), the previously claimed evidence for point sources as the bulk of the excess disappears. That is, the 2015 wavelet result in Ref.~\cite{Bartels:2015aea} was correct in that it detected previously unknown point sources, but now such point sources are known to be mostly part of Fermi's 4FGL point source catalog, which cannot be the bulk of the excess. This is because masking the 4FGL catalog does not change the intensity of the excess, while stacking the spectra of non-excluded potential GCE sources reveals that the non-excluded sources do not produce enough flux to power the excess~\cite{Zhong:2019ycb}. This points towards a \textit{smooth} origin for the excess, rather than a bright point source origin. Any remaining point sources, if they exist, must be too faint to be seen, and being so faint, many more of them are required to produce enough flux to explain the excess. This places the number of point sources required in the several to tens of thousands, or potentially up to a few million sources depending on the cutoff on the low flux end of the luminosity function~\cite{Zhong:2019ycb}. The upper ends of these estimates are certainly in strong tension with the maximum number of pulsars that could possibly exist in the inner galaxy. In future, better understanding of the total number of pulsars in the Milky Way may set a strong constraint on the potential luminosity function of a new population of pulsars needed to explain the excess.\\

\textbf{2020 Updates}\\

In Ref.~\cite{Buschmann:2020adf}, it was found that the failure of the injection test reported in Ref.~\cite{Leane:2019xiy} was indeed likely due to diffuse mismodeling. This is because simulating a dataset using a newer and improved Galactic diffuse emission model than those used previously, and fitting the data with the older Galactic diffuse emission models, could replicate the negative dark matter fluxes found when fitting the real data (in the scenario the known point sources are masked). In addition, a new way to improve the Galactic diffuse emission model was presented, by breaking the Galactic diffuse emission model into spherical harmonics and floating the pieces separately. In that case, in a 25 degree radius ROI (region of interest) in the sky with a $\pm2$ degree band mask, it was found that a GCE point source flux was found in all Galactic diffuse emission models. However, compared to the original NPTF paper~\cite{Lee:2015fea}, the point source evidence in this analysis with the new model is now decreased to only around $\sim3\sigma$. This evidence drops even further for any other radius choice for the ROI other than the 25 degree radius ROI shown in Ref.~\cite{Buschmann:2020adf}. 

Improving the Galactic diffuse emission model, as done in Ref.~\cite{Buschmann:2020adf}, is an important step forward, as this is currently the least well-understood component of the gamma-ray sky. Indeed, even the best models we currently have, including those used in Ref.~\cite{Buschmann:2020adf}, still lead to very poor fits to the data. This means there is already another systematic \textit{we know} is present and that \textit{we do not know} the impact of: what happens to the result once there is a correct Galactic diffuse emission model (one that actually matches the data to the level of Poisson noise). As such, it is not clear if systematic effects have yet been fully bracketed.

Indeed, the importance of unidentified systematics, and the implications for attempting to claim any evidence for point sources in the data, were further argued recently in Refs.~\cite{Leane:2020nmi,Leane:2020pfc}.
There, it was explicitly shown that mismodeling of a smooth signal, and a true point source signal, both lead to increased pixel-to-pixel variance in the data, and that the fit can mistake one for the other. This means that apparent evidence for a point source signal can be misinterpreted, when truly its origin may just be in mismodeling. This argument was mathematically demonstrated in Ref.~\cite{Leane:2020pfc}. Given we already know we are not modeling the gamma-ray sky perfectly, this is a serious concern for claimed evidence using the NPTF. 

Strikingly, evidence of this effect has been found in analyses of real data. In a 10 degree ROI (the region where the GCE is brightest), it was shown that the apparent evidence for point sources using the real Fermi gamma-ray data can be instead be directly traced to manifestations of mismodeling, and lack of inclusion of important systematics~\cite{Leane:2020nmi,Leane:2020pfc}. It was demonstrated, in the real data and this ROI, that simply allowing the GCE template to float freely in north and south pieces results in an asymmetric smooth excess, and makes the point source evidence disappear. This behavior was reproduced in simulations, which revealed using an overly restrictive model template created a spurious point source population. This has a source count function that is consistent with what is found in all existing NPTF GCE studies of real data, including Refs.~\cite{Lee:2015fea,Buschmann:2020adf}. Indeed, this is suggestive (but does not prove) that potentially all NPTF studies are finding a spurious point source population as a result of some kind of mismodeling. The asymmetry of the GCE found in Refs.~\cite{Leane:2020nmi,Leane:2020pfc} is likely a manifestation of \textit{more} unknown systematics, likely transferred from mismodeling the Galactic diffuse emission model\footnote{Note that the GCE smooth asymmetry appears using the latest Galactic diffuse emission model in Ref.~\cite{Buschmann:2020adf}, as well.}. As such, the asymmetry is not argued to be an intrinsic feature of the excess itself. This is because the degree of asymmetry appears to depend on the Galactic diffuse emission model used and ROI. However, if it were shown to actually be the true distribution of the excess, it would strongly disfavor a dark matter interpretation (though it is also not clear how a population of point sources would exist with such a large asymmetry).

The original GCE NPTF study~\cite{Lee:2015fea} noted that while unaccounted for systematics could potentially affect the reported results, the results appeared robust to the range of systematics tested.  Refs.~\cite{Leane:2019xiy,Leane:2020nmi,Leane:2020pfc}, on the other hand,  identified large systematic effects which change the interpretation substantially, explicitly showing what can (and does) go wrong when systematics are missed. While improvements certainly could be made in future, given the impact of systematics at the moment, it does not appear that a dominantly smooth GCE (or point source GCE) is in tension with NPTF analyses. At this stage, we just don't know what the answer is!\\

\textbf{Finally settling the debate}\\

Within the next decade, it should likely be possible to answer the question of the origin of the excess. Steps that can be taken include:

\begin{itemize}
 \item New measurements with an MeV gamma-ray instrument could allow the low-energy part of the GCE energy spectrum to be measured more precisely. This is because the systematics using the Fermi telescope degrade drastically in the sub-GeV energy band. In this range, Fermi's PSF is a few degrees, while MeV-targeted instruments may allow for the sub-1 degree resolution. Measuring the low-energy part of the spectrum is important as this is where the DM and pulsar energy spectra predictions deviate most substantially.
 \item Discovery of more dwarf spheroidal galaxies, to increase sensitivity of Fermi to the dark matter annihilation cross section (see sub-section below).
 \item Observing the candidate pulsars directly in radio, if they exist (see radio waves section below).
 \item New measurements of the local dark matter density with Gaia. Currently, systematics on the dark matter density profile are a significant source of uncertainty, which when tightened will help corner the dark matter explanation.
  \item New and better Galactic diffuse emission models. This is key on the theory front, and is certainly the most significant barrier at the moment to obtaining accurate models of the sky. This can be achieved by better understanding cosmic ray propagation, or obtaining better dust maps with Planck. (The dust provides a tracer of the gas distribution.)
\end{itemize}

\subsubsection{Dark Matter Annihilation Limits}

Gamma rays from the center of the galaxy can also be used to set limits on dark matter annihilation. 

\begin{figure}[t!]
\centering
\includegraphics[width=0.45\textwidth]{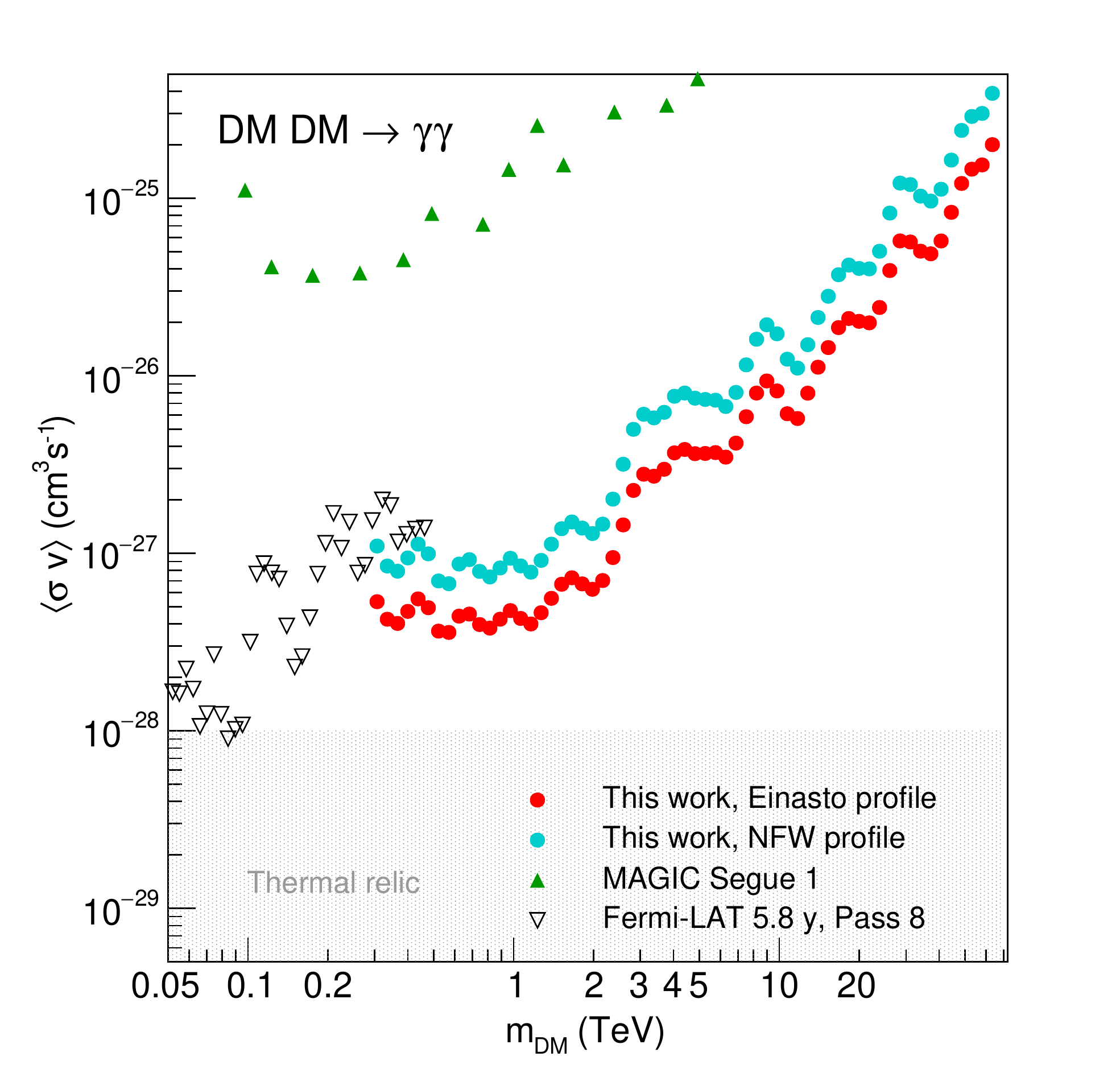}
\hspace{2mm}
\includegraphics[width=0.46\textwidth]{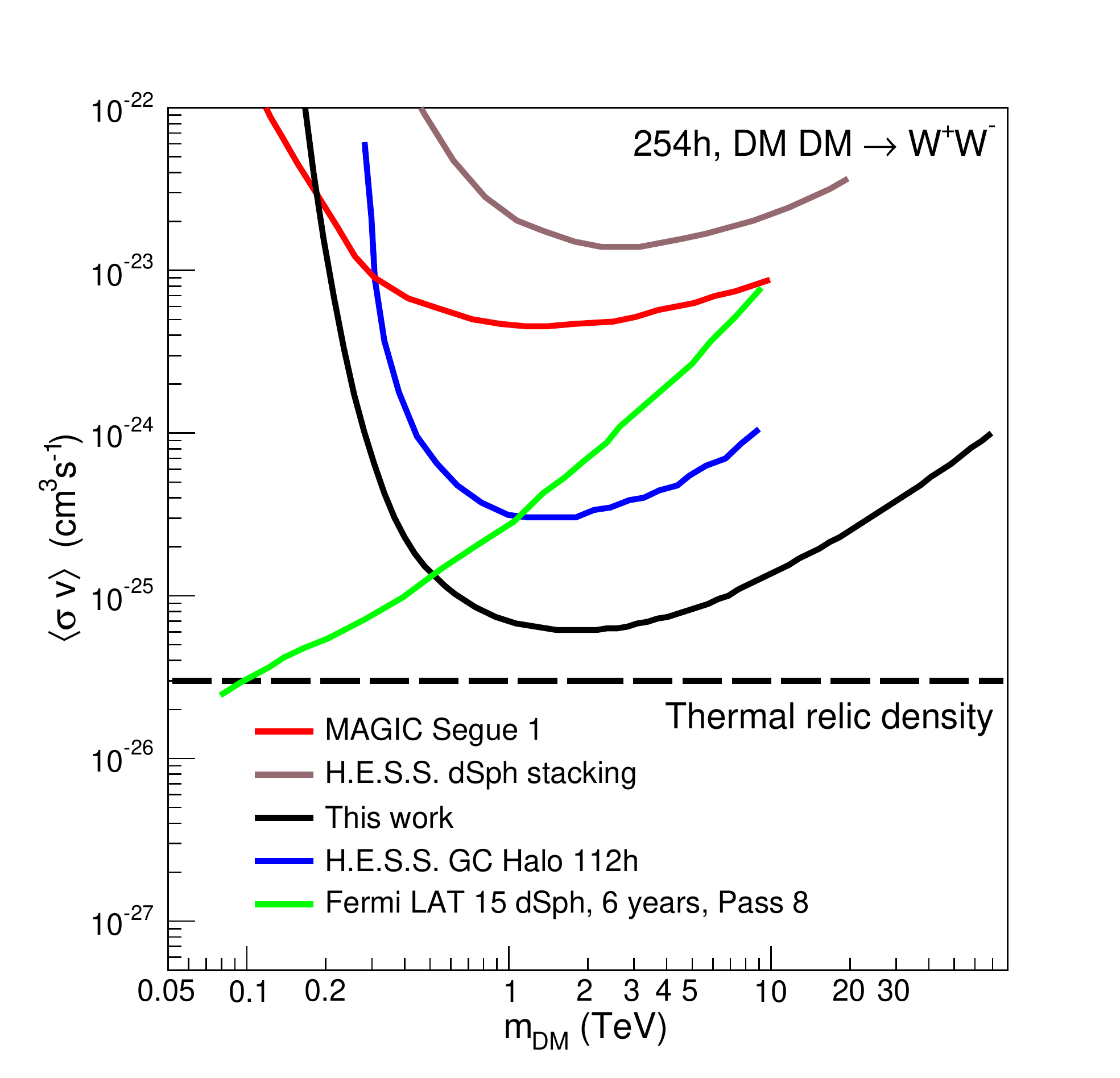}
\caption{\textbf{Left:} Galactic center gamma-ray line limits from H.E.S.S. (with comparison shown to Fermi and MAGIC), figure taken from Ref.~\cite{Abdallah:2018qtu}. \textbf{Right:} Annihilation to WW limits from H.E.S.S., also with comparison with Fermi and MAGIC. Figure taken from Ref.~\cite{Lefranc:2016srp}, where other final state limits can also be found.}
\label{fig:hess}
\end{figure}

Figure~\ref{fig:hess} shows bounds on TeV dark matter annihilation in the inner galaxy, from both line searches (direct annihilation into gamma-ray pairs), as well as other various final states. The limits shown are from H.E.S.S. as per Ref~\cite{Abdallah:2018qtu} and Ref~\cite{Abramowski:2011hc,Lefranc:2016srp}, as these are the strongest bounds in the TeV region (though other experiments, Fermi, MAGIC, are compared in the figures). Note that while not shown in the figure, there are also Galactic halo bounds from HAWC~\cite{Abeysekara:2017jxs}). For decay, see e.g. Refs.~\cite{Pierre:2014tra,Abeysekara:2017jxs} (and for leading gamma-ray limits on decay from other environments see e.g. Refs.~\cite{Essig:2013goa,Cohen:2016uyg}).

In the GeV region, if a bulge morphology rather than dark matter is assumed for the GCE, strong limits can be set on the dark matter annihilation~\cite{Abazajian:2020tww}.

\subsection{Dwarf Spheroidal Galaxies}

The dark matter dense satellites of the Milky Way are called dwarf spheroidal galaxies. Having a high signal to background ratio, they are ideal targets for dark matter searches. They yield the strongest limits on dark matter annihilation to any final states which produce copious gamma rays, i.e. hadrons. In the GeV mass range, limits from Fermi are superior. While $\sim2\sigma$ local excesses have been observed in some dwarfs~\cite{Fermi-LAT:2016uux}, this is not globally significant at this time.

Figure~\ref{fig:limits} (middle) shows limits from Fermi observations of gamma rays from dwarf spheroidal galaxies, for various annihilation products. See Ref.~\cite{Fermi-LAT:2016uux} for the most recent official Fermi analysis.

More recently, the impact of systematics on dwarf results has been demonstrated to weaken bounds by almost an order of magnitude~\cite{Ando:2020yyk}. The largest source of uncertainty here is the dark matter density profiles, which are modeled based on N-body simulations. This leaves no tension at the moment with dark matter explanations of the GCE.

Moving forward, dwarf spheroidal galaxies are a leading target to corroborate a potential dark matter signal at the galactic center. The Dark Energy Survey (DES) and LSST are expected to locate more dwarf galaxies, which can increase potential sensitivity to the dark matter annihilation cross section by more than an order of magnitude~\cite{Charles:2016pgz}. This may lead to an exclusion, or a corroborated signal with the galactic center. 

\begin{figure}[t]
\centering
\includegraphics[width=0.52\textwidth]{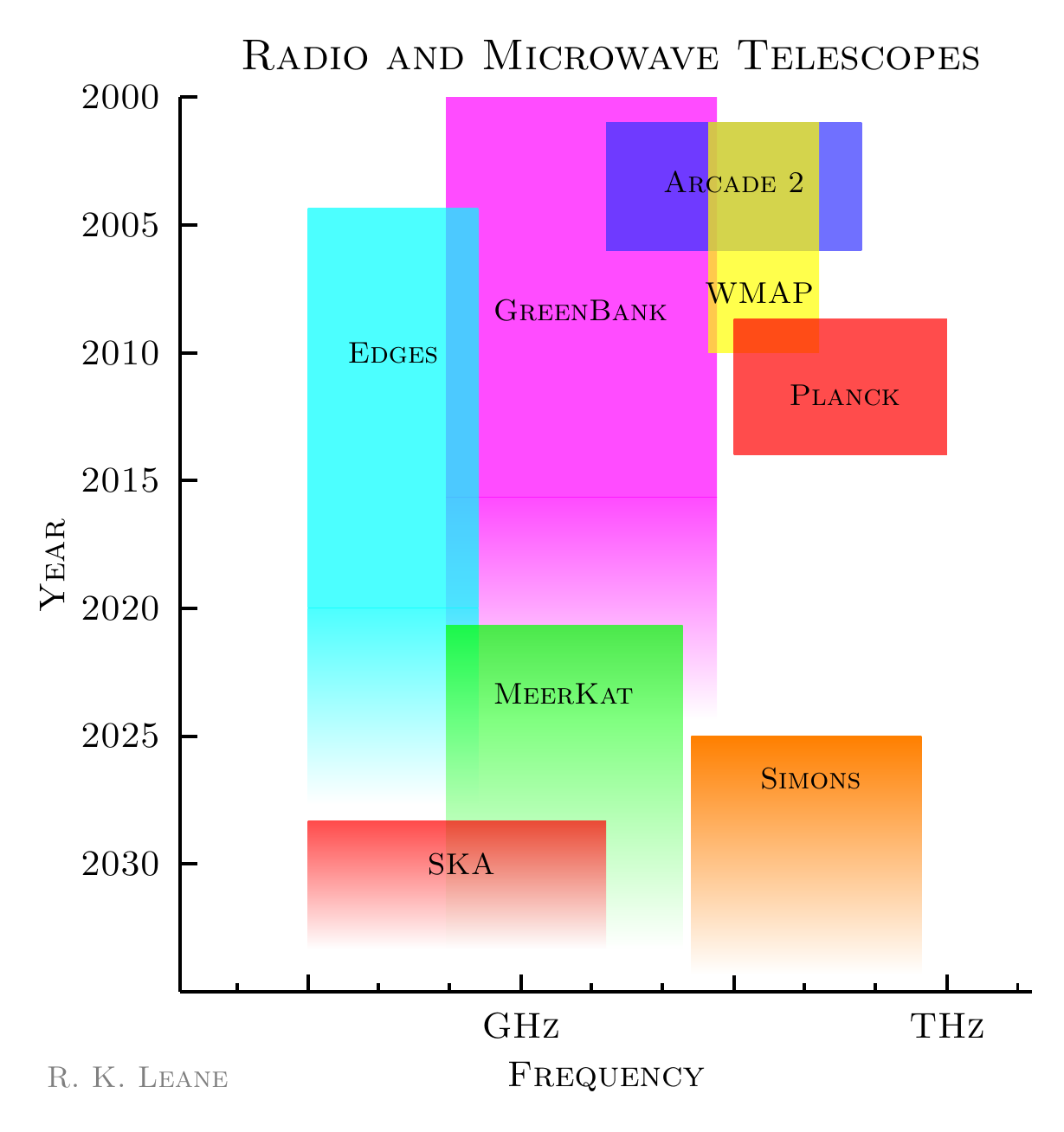}
\includegraphics[width=0.47\textwidth]{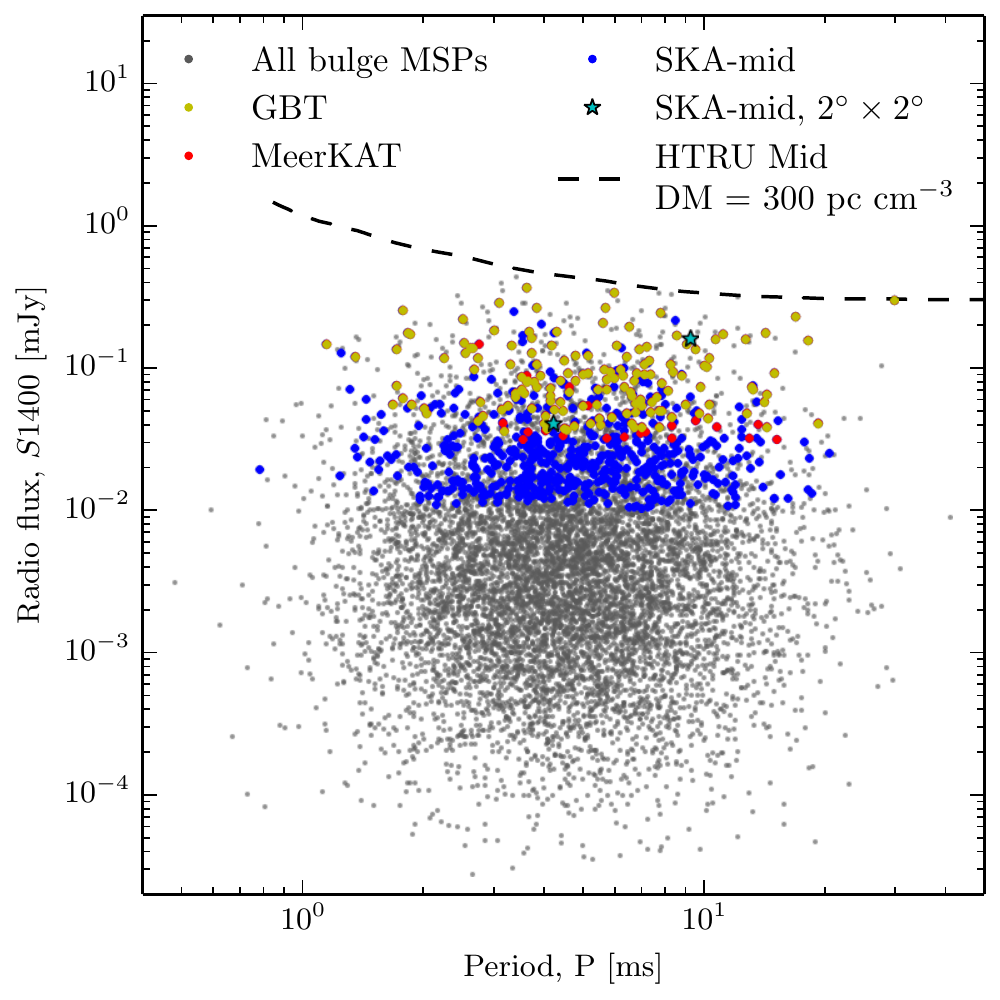}
\caption{\textbf{Left:} Summary of selected micro and radiowave experiments by year operating, and approximate frequency sensitivity. \textbf{Right:} Discovery potential for GCE millisecond pulsars in radio. Right figure taken from Ref.~\cite{Calore:2015bsx}.}
\label{fig:radiomicro}
\end{figure}

\section{Dark Matter Searches with Micro and Radio Waves}

\subsection{Experiments and Prospects}

Dark matter annihilation can lead to radio waves, by their charged annihilation products producing synchrotron emission as they pass through astrophysical magnetic fields (see e.g.~Refs.\cite{Colafrancesco:2005ji,Fornengo:2011xk}).

Radio waves are also interesting because they can be used to search for dark matter impostors -- pulsars. In terms of the GCE, pulsars potentially present in the galactic center may pulse into radio as well as gamma rays. If so, these could be detected by the upcoming observations of \textsc{MeerKat} or the SKA~\cite{Calore:2015bsx}.

Figure~\ref{fig:radiomicro} shows these prospects. On the left, selected radio and micro wave telescopes are shown, with their approximate frequency sensitivities and mission times. On the right, the prospects to find galactic center pulsars in radio are shown. In principle the GreenBank telescope may already be able to see some of these pulsars if they exist, however no studies have been reported so far.

Dark matter annihilation or decay can also lead to observables in microwaves. Of particular interest is the impact on the cosmic microwave background. Upon annihilating or decaying during the recombination era, charged annihilation products can inject ionizing energy which alters the ionization history of the Universe, and expected features of the CMB~\cite{Slatyer:2009yq,Slatyer:2015jla}. This provides the strongest probe of dark matter annihilation to visible products below about 10 GeV, as shown in Fig.~\ref{fig:limits} (left). CMB measures of dark matter annihilation are not expected to drastically improve with future measurements, simply because we are cosmic variance limited (we only have one Universe to measure!). However, best improvements can be expected with the upcoming Simons Observatory.

\section{Dark Matter Searches with Charged Cosmic Rays}

\subsection{Experiments and Prospects}

For dark matter annihilation into leptonic final states, the best sensitivity is usually obtained by studying charged cosmic rays. One of the exciting aspects of cosmic-ray research is the extreme energies available to probe new physics, which far exceed anything ever made on Earth. We have high-energy cosmic-collider beams just waiting to be studied in new ways. A difficulty in this search is that cosmic ray propagation is not well understood and often induces substantial systematic uncertainties.

\begin{figure}[t!]
\centering
\includegraphics[width=0.9\textwidth]{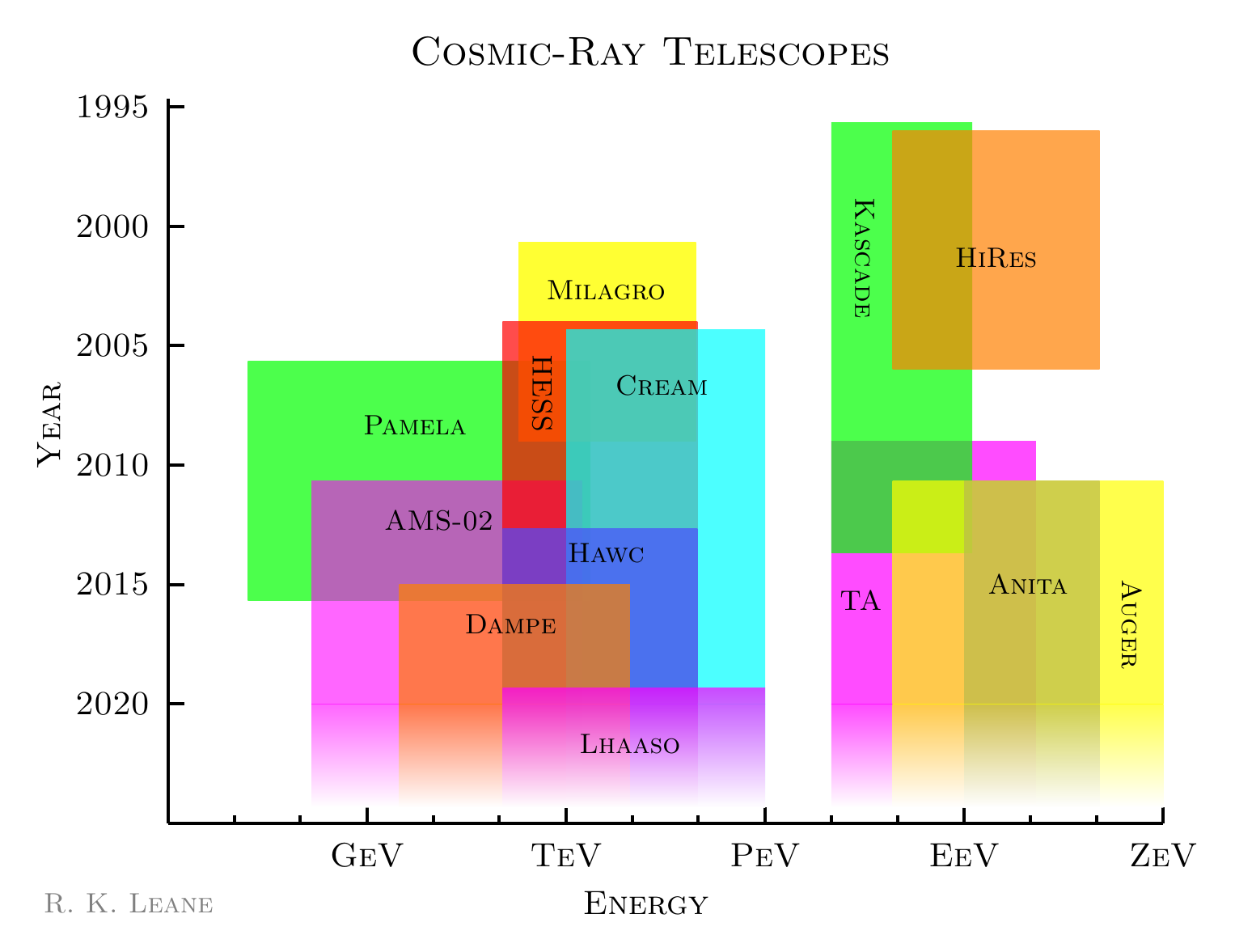}
\caption{Summary of selected cosmic-ray experiments by year operating, and approximate energy sensitivity. Balloon/transient experiments (e.g. Cream and ANITA) which have discrete missions have been shown as continuous over the time the collaboration is still active.}
\label{fig:cosmicrays}
\end{figure}

Figure~\ref{fig:cosmicrays} summarizes cosmic-ray experiments searching for dark matter interactions, as a function of their mission dates.

\subsection{Positrons}

An excess in $\sim10-1000$ GeV positrons has been reported by PAMELA~\cite{Adriani:2008zr}, AMS-02~\cite{Aguilar:2013qda,Accardo:2014lma}, and recently DAMPE~\cite{Ambrosi:2017wek}. If fitting to annihilating dark matter, it is consistent with a mass of $\sim$TeV. The annihilation cross section for this process, however, is $\sim10^{-23}$cm$^3$/s, which is 3 orders of magnitude above that expected for a vanilla thermal dark matter relic. This excess sparked increased interest in Sommerfeld enhanced dark matter annihilation~\cite{ArkaniHamed:2008qn}, where the rate of annihilation is much larger in the Universe today than it was at freeze out, due to enhancement from long-range forces. It also sparked interest in leptophilic dark matter models (see e.g.~\cite{Fox:2008kb,Kopp:2009et,Bell:2014tta, DEramo:2017zqw}), to order to produce the excess with a large branching fraction to leptons, and avoid other hadronic constraints.

This signal was presented also as likely coming from pulsars~\cite{Hooper:2008kg,Hooper:2017gtd}, as they are expected to produce a comparable positron energy spectrum as that observed in the positron excess. An argument against this result was reported by the HAWC collaboration in 2017~\cite{Abeysekara:2017old}, claiming that the excess could \textit{not} be due to pulsars, as the diffusion coefficient observed near two of the closest pulsars to us, Geminga and Monogem, appeared to be too low to allow their positrons to escape to reach us at Earth (in fact, it was measured to be about two orders of magnitude lower than expected elsewhere in the Galaxy). However, globally having such a low diffusion coefficient would mean that cosmic rays \textit{in general} would struggle to reach us -- and we know that cosmic rays do reach us on Earth, given that we have detected them! So, we know they mustn't be slowed forever by low diffusion. The conclusion to this saga is that the diffusion coefficient is not uniform throughout the Galaxy~\cite{Hooper:2017tkg,Profumo:2018fmz}, and they can reach us from these pulsars. Therefore, these pulsars remain the most likely explanation of the positron excess at this stage, rather than annihilating dark matter.

\begin{figure}[t]
\centering
\includegraphics[width=0.325\textwidth]{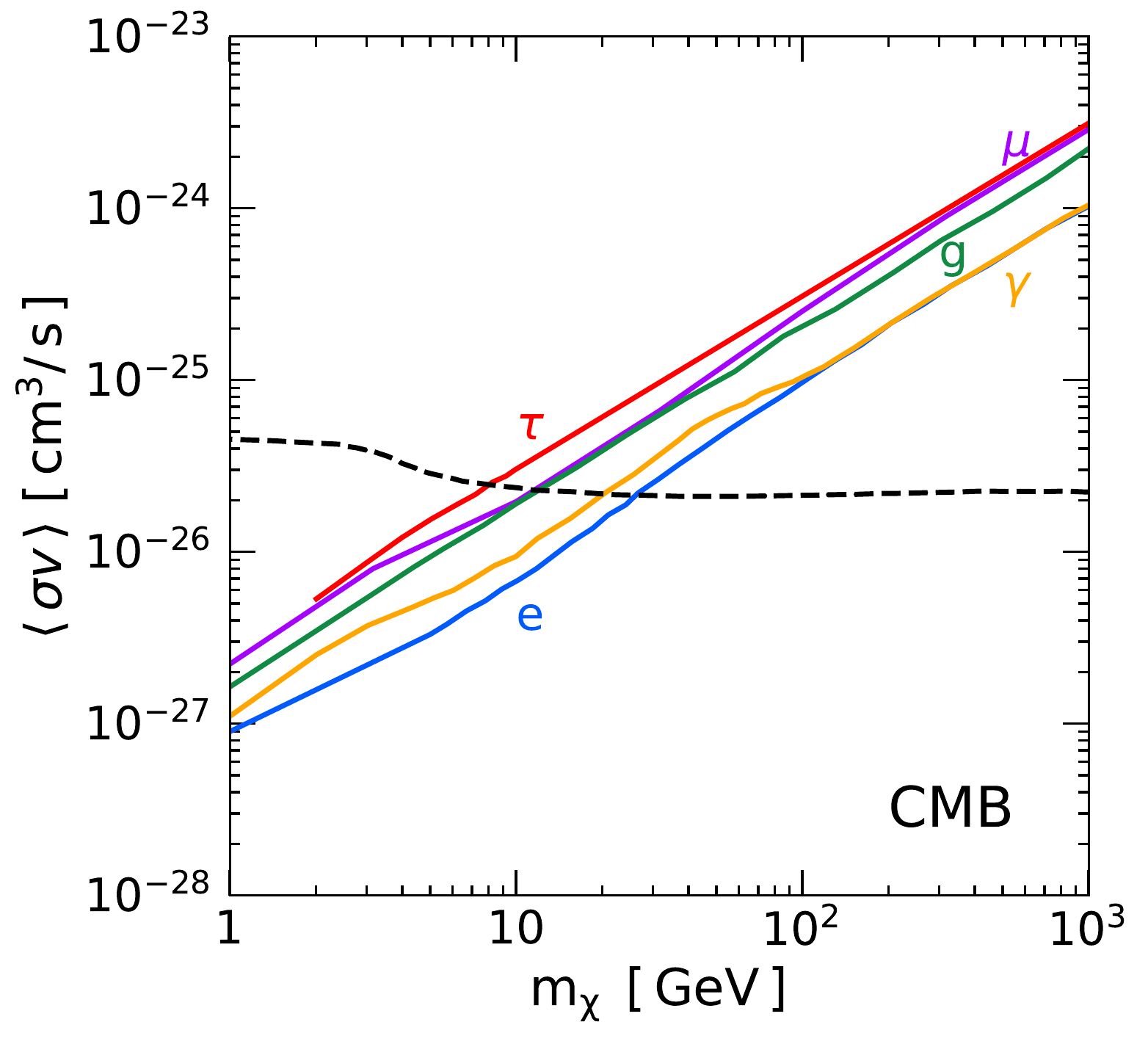}
\includegraphics[width=0.325\textwidth]{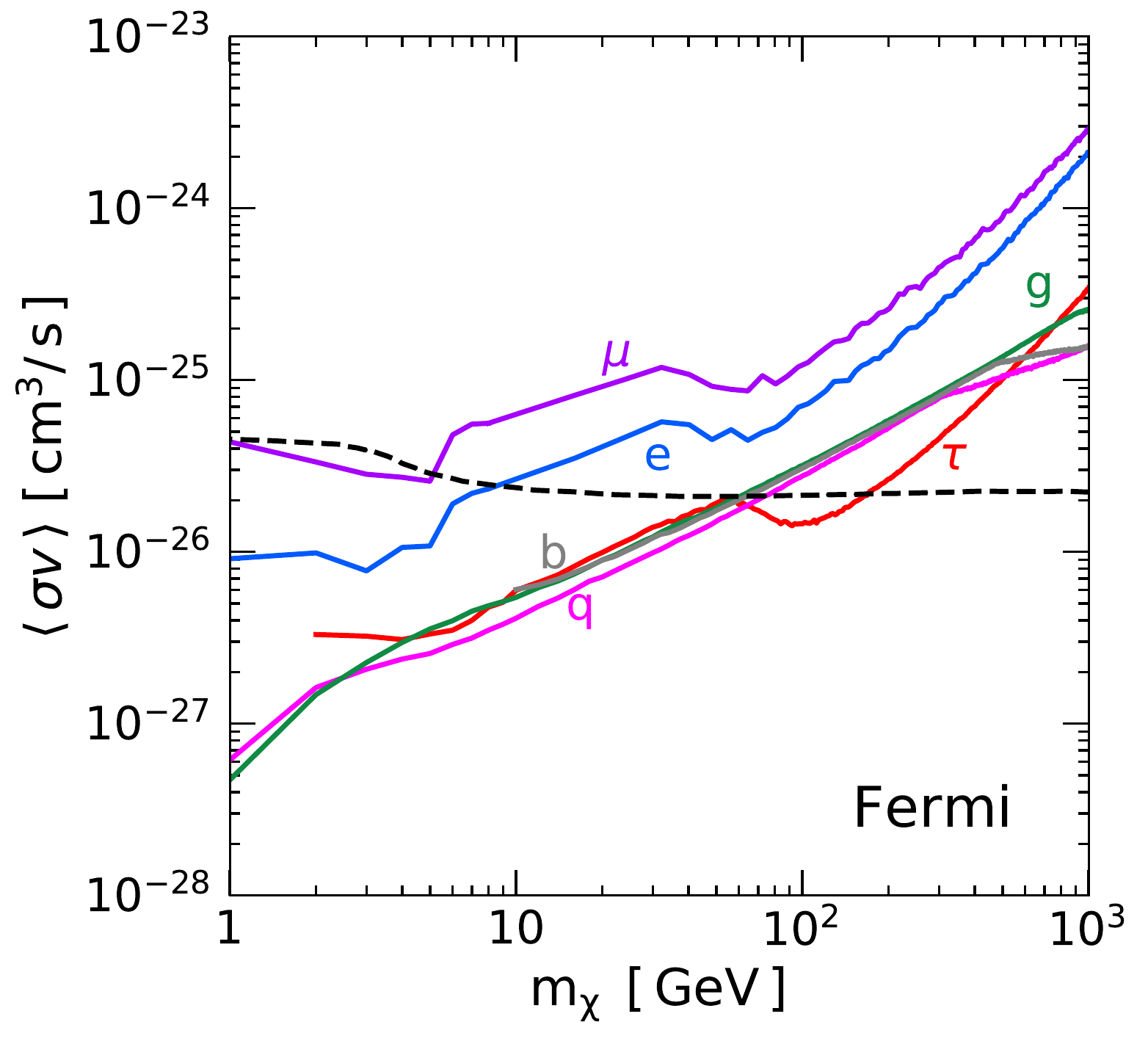}
\includegraphics[width=0.325\textwidth]{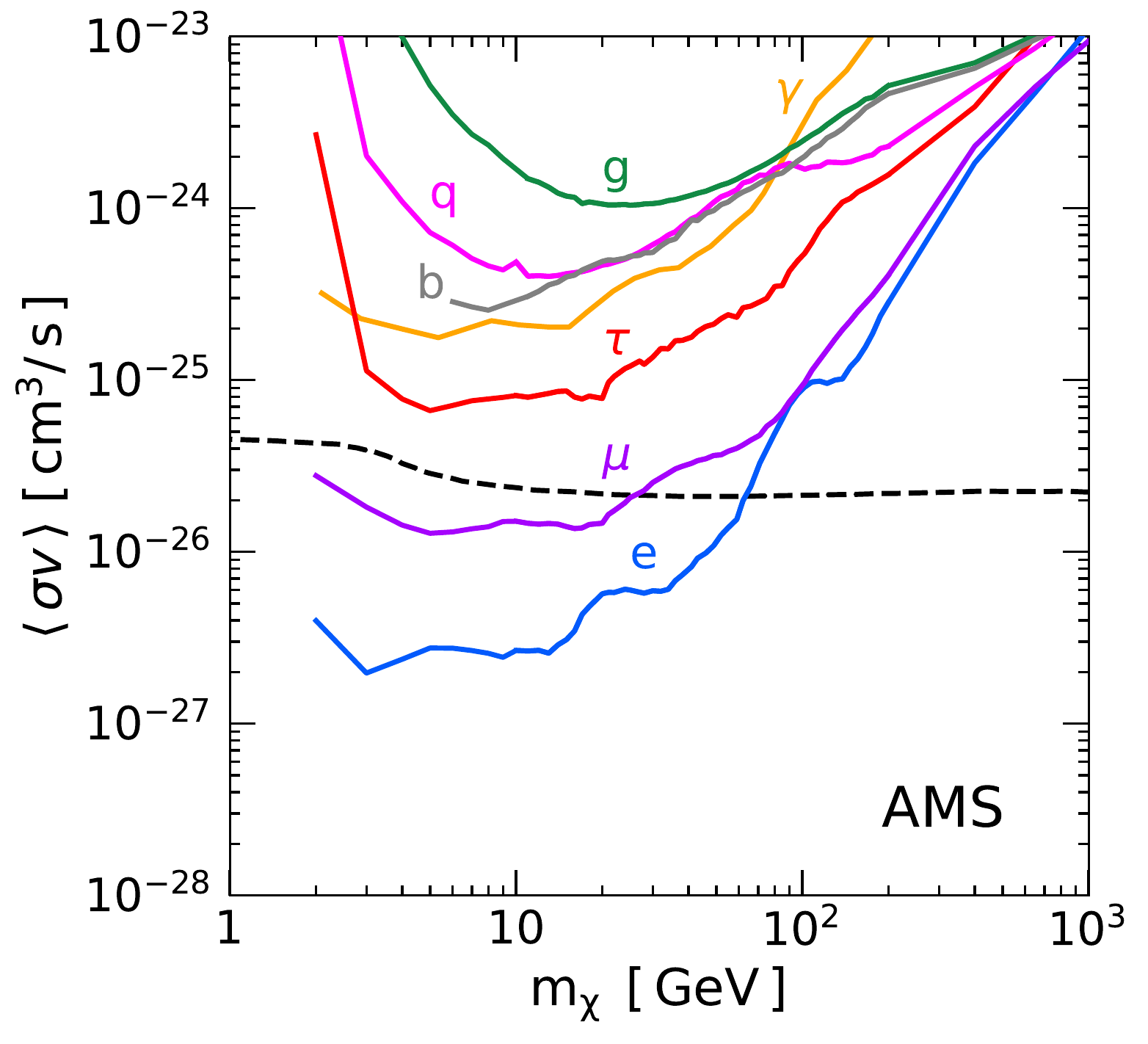}
\caption{Conservative limits on $s$-wave $2\rightarrow2$ GeV dark matter annihilation to various final status as labeled. {\bf Left:} Limits from the CMB. {\bf Middle:} Limits arising from Fermi measurements of gamma rays from dwarf spheriodal galaxies. {\bf Right:} Limits from positron flux measurements with AMS. Relic cross section is the dashed line~\cite{Steigman:2012nb}. Figures taken from Ref.~\cite{Leane:2018kjk}.}
\label{fig:limits}
\end{figure}

For limits on annihilating DM from positrons from AMS, see Fig.~\ref{fig:limits} (right). 

\subsection{Antiprotons}

AMS-02 has reported an excess of antiprotons with energies $\sim5-10$ GeV~\cite{Aguilar:2016kjl}. This can be well fit with an annihilating dark matter candidate of mass $\sim50-100$ GeV~\cite{Cuoco:2019kuu,Cholis:2019ejx}, particularly if it annihilates to hadronic final states. This excess becomes even more intriguing when noting it is consistent with a common origin with the GCE, as well as the thermal relic cross section, with intensity $\langle\sigma v \rangle\sim10^{-26}$cm$^3$/s. Both excesses can be fit in a mass range of $50-70$ GeV assuming $2\rightarrow2$ $s$-wave annihilation to $b$-quarks~\cite{Cuoco:2017rxb,Cholis:2019ejx}. In the scenario dark matter exists in a hidden sector, there are a number of well motivated models that can fit both excesses, and have zero tension with collider, direct detection, dwarf or relic density constraints~\cite{Hooper:2019xss}.

To understand the origin of this excess, it would be ideal if AMS would release their uncertainty correlation matrices. As these are not public, educated guesses need to be made as to the correlated uncertainties between energy bins. One group has argued that the significance of the signal ranges from $3.3-7.7\sigma$ depending on cosmic-ray propagation values~\cite{Cholis:2019ejx}, while another argues it is $\sim3-5.5\sigma$~\cite{Cuoco:2019kuu}. On the other hand, it has been shown that this excess might just be consistent with cosmic-ray secondaries (i.e. not dark matter)~\cite{Boudaud:2019efq,Heisig:2020nse}. Indeed, while all these antiproton studies do consider the sets of systematics, Refs.~\cite{Boudaud:2019efq,Heisig:2020nse} argue that they should be significantly larger than previously estimated, which is what reduces their calculated significance of the excess. Moving forward, better understanding cosmic-ray propagation will be key to interpret signals as they arise. 

For antiproton constraints on dark matter annihilation, see e.g. Refs.~\cite{Giesen:2015ufa,Cuoco:2016eej,Cuoco:2017iax}.
\\

\begin{figure}[h!]
\centering
\includegraphics[width=0.5\textwidth]{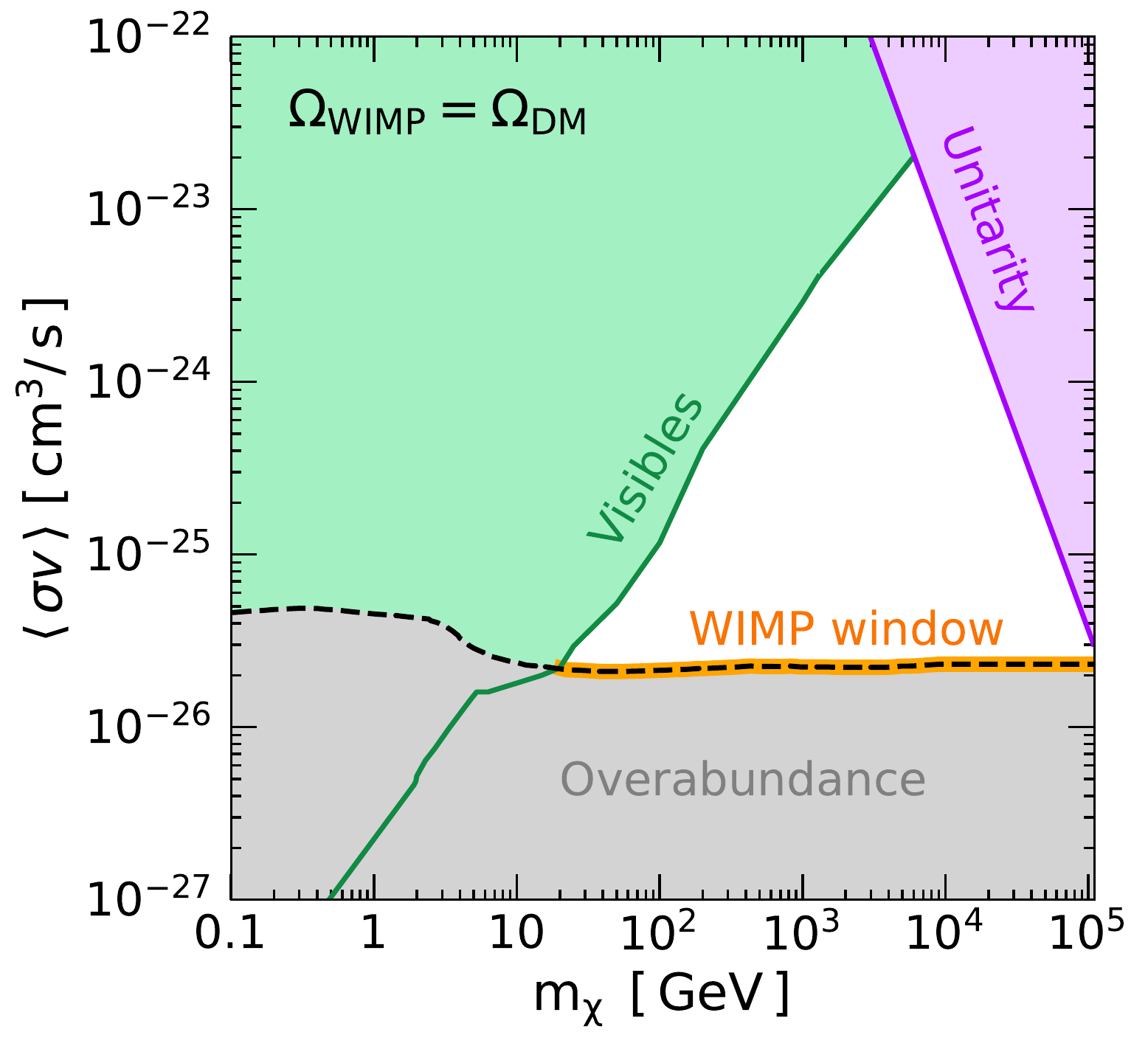}
\caption{Combined lower limit on the dark matter cross section, for $2\rightarrow2$ $s$-wave annihilation to visible final states. Dashed line is thermal relic cross section. Figure taken from Ref.~\cite{Leane:2018kjk}.}
\label{fig:window}
\end{figure}

\section{Outlook}

After all this, we can ask, what is the status of indirect dark matter searches? 

Figure~\ref{fig:limits} summarizes the strongest and most robust bounds for GeV dark matter annihilating into various final states; Fermi observations of dwarf spheroidal galaxies provide the strongest bounds on annihilation to hadronic final states, AMS-02 positron fluxes restrict most annihilation to leptons, and the CMB limits are strongest for low dark matter masses.

Figure~\ref{fig:window} shows what the limit looks like if we combine \textit{all} this complementary information for each type of visible final state. If one experiment constrains annihilation into a particular final state, the remaining energy at a given mass must go into a different final state, which can be ruled out by a complementary experiment. If no final states can proceed without exclusion, that particular mass is considered ruled out. This plot shows how much progress we have really made in probing the thermal WIMP. It looks like not much so far -- just up to 20 GeV! The window up to the unitarity limit of $\sim$100 TeV~\cite{Griest:1989wd,Smirnov:2019ngs} remains largely open. This window can only be closed by indirect detection experiments, which directly probe the annihilation rate. This motivates many interesting searches and improvements for years to come.

Indeed, indirect detection searches for dark matter are exciting now more than ever. We are exploring unprecedented parameter space, and have a number of persistent hints which could potentially turn out to be dark matter signals. 

In the center of our galaxy, the excess of GeV gamma rays reported by Fermi-LAT might be dark matter. The leading alternate explanation is millisecond pulsars. In terms of the non-Poissonian template fitting analyses, as systematics have been shown to not be fully controlled, evidence does not clearly support a smooth nor point source origin at this stage. Recent wavelet analyses have shown that the previous apparent point source detection cannot make up the bulk of the GCE. Dwarf limits have, also due to systematics, recently been shown to be potentially weaker than expected, leaving open the masses and annihilation cross sections that could explain the GCE. Measurements of the excess morphology are actively improving, as recent studies have shown that the morphology may be more like the stellar bulge than a dark matter profile, which would support the non-dark matter interpretation. However, given all models are currently poor fits to the gamma-ray data, it is difficult to know if systematics are truly bracketing all the uncertainties they should be. This will certainly remain an interesting open problem for the next few years, and has many observational prospects on the horizon to help settle the debate.

Antiprotons are observed in excess at AMS-02, and while it is consistent with a dark matter origin (and consistent with the GCE), there are also arguments that it is consistent with cosmic-ray secondaries. Systematic correlation matrices, and better cosmic-ray propagation models, are needed here to definitively confirm or exclude explanations of this excess.

Positrons exist in excess, as observed by PAMELA, AMS-02, and DAMPE. This signal is consistent with a $\sim$TeV dark matter candidate. However, the current consensus is that this is likely due to pulsars, due to nearby known pulsars Geminga and Monogem presenting a comparable positron energy spectrum. Regardless, debate over this signal has led to very interesting developments in understanding cosmic-ray propagation. In particular, that the diffusion coefficient is not uniform throughout the Galaxy.

Lastly, the 3.5 keV X-ray line remains the subject of lively debate, as it might be evidence of the decay of a $\sim$7 keV sterile neutrino dark matter candidate. There are arguments it is ruled out from blank-sky observations, and counter-arguments that it is not, due to particular systematics and modeling issues. Upcoming observations with new X-ray telescopes should answer the question once and for all.

The interplay of all these observations, and corroboration of potential dark matter signals as they arise, will be crucial to understand the nature of dark matter in the years to come.

\section*{Acknowledgments}

I thank the organizers of the EDSU conference for the invitation and an excellent semi-virtual conference, and John Beacom and Tracy Slatyer for comments on the manuscript. I acknowledge support from the Office of High Energy Physics of the U.S. Department of Energy under Grant No. DE-SC00012567 and DE-SC0013999, as well as the NASA Fermi Guest Investigator Program under Grant No. 80NSSC19K1515.

\bibliographystyle{JHEP}
\bibliography{references}

\end{document}